\documentclass[final, times]{elsarticle}

\usepackage{fancyhdr}
\fancyheadoffset{58pt}
\fancypagestyle{fancytitle}{
\fancyfoot{}
\lfoot{\small \textit{This is a manuscript version of an article published in Computers and Electronics in Agriculture, available at:} \url{https://doi.org/10.1016/j.compag.2021.106183}}
}

\usepackage{lineno,hyperref}
\modulolinenumbers[5]

\usepackage{multirow}
\usepackage{booktabs}

\graphicspath{{figures/}{./}}

\journal{Computers and Electronics in Agriculture}

\begin{document}
\begin{frontmatter}
\title{AHMoSe: A Knowledge-Based Visual Support System for Selecting Regression Machine Learning Models}

\author[KUaddress]{Diego Rojo\corref{mycorrespondingauthor}}
\cortext[mycorrespondingauthor]{Corresponding author}
\ead{diego.rojogarcia@kuleuven.be}

\author[KUaddress]{Nyi Nyi Htun}
\ead{nyinyi.htun@kuleuven.be}

\author[PUCaddress,MIFDaddress]{Denis Parra}
\ead{dparra@ing.puc.cl}

\author[KUaddress]{Robin De Croon}
\ead{robin.decroon@kuleuven.be}

\author[KUaddress]{Katrien Verbert}
\ead{katrien.verbert@kuleuven.be}

\address[KUaddress]{Department of Computer Science, KU Leuven, Leuven, Belgium}
\address[PUCaddress]{Department of Computer Science, Pontificia Universidad Catolica de Chile, Santiago, Chile}
\address[MIFDaddress]{Millennium Institute Foundational Research on Data, Santiago, Chile}

\begin{abstract}
Decision support systems have become increasingly popular in the domain of agriculture. With the development of automated machine learning, agricultural experts are now able to train, evaluate and make predictions using cutting edge machine learning (ML) models without the need for much ML knowledge. Although this automated approach has led to successful results in many scenarios, in certain cases (e.g., when few labeled datasets are available) choosing among different models with similar performance metrics is a difficult task. Furthermore, these systems do not commonly allow users to incorporate their domain knowledge that could facilitate the task of model selection, and to gain insight into the prediction system for eventual decision making. 
To address these issues, in this paper we present AHMoSe, a visual support system that allows domain experts to better understand, diagnose and compare different regression models, primarily by enriching model-agnostic explanations with domain knowledge. 
To validate AHMoSe, we describe a use case scenario in the viticulture domain, grape quality prediction, where the system enables users to diagnose and select prediction models that perform better. We also discuss feedback concerning the design of the tool from both ML and viticulture experts.
\end{abstract}


\end{frontmatter}
 \thispagestyle{fancytitle}


\section{Introduction}
Decision support systems (DSSs) are often used in places that call for domain experts to deal with massive amounts of data. In the field of agriculture, for example, various stakeholders such as farmers, advisers and policymakers tend to use DSSs to facilitate farm management and planning tasks~\cite{gutierrez2019review}. Typically, data is first gathered from multiple sources including sensors, satellites and in-field observations, and analysed using a series of statistical models. The output of such an analysis is finally used for the required decision making task.

Recent popularity and success of machine learning (ML) has also led to an increased use of off-the-shelf ML systems that can be used by domain experts with little to no machine learning knowledge. Examples include agricultural experts employing a variety of graphical user interfaces (GUIs) to predict yield, revenue, resource requirements, etc.~\cite{gutierrez2019review}. To support such use cases, the field of automated machine learning (AutoML) has elaborated several solutions that automate the ML pipeline such as feature engineering~\cite{Kaul2017, Katz2016}, hyperparameter optimisation~\cite{Feurer2019} or neural networks architecture search~\cite{Elsken2019}. After receiving an input dataset, an AutoML system builds and evaluates (usually hundreds of) different prediction models, presenting the user with a leader board with the models ordered by their performance score. Several AutoML products such as H2O Driverless AI~\cite{hall2019} provide a GUI to enable end-users to input a dataset and configure parameters of the AutoML process, such as the time or the model's families to consider. All these advances have empowered non-technical users to train and make predictions using cutting edge machine learning models. 

Unfortunately, there are still several challenges to using these AutoML systems. One such challenge is related to the interpretation and selection of models that have very similar performance scores. A good heuristic is choosing the simplest model among those models performing similarly well, for instance the model with fewer parameters~\cite{domingos1999role}. However, even this task can be difficult if the simplest model is not aligned with experts' knowledge, i.e., if the features that contribute the most to the model predictive performance are not aligned with experts' knowledge. Gil et al.~\cite{Gil2019} suggested that the use of supplementary information extracted from domain knowledge of the experts is a potential strategy to evaluate and select a model. Such supplementary information includes rules defined by domain experts for knowledge-based DSSs (e.g., fuzzy inference systems~\cite{Papadopoulus2011}). Enabling users to incorporate domain knowledge into the model assessment and selection process of AutoML systems could help diagnose models that struggle to generalise and identify the ones that agree with domain expert knowledge.

Another challenge of using AutoML as DSSs is their inherent lack of transparency. The models employed to date for agricultural DSSs~\cite{accorsi2014hydroqual, han2017climate, rossi2014addressing, tayyebi2016smartscape} remain opaque to users and enclosed behind the software. Hence, they are often referred to as ``black-boxes''~\cite{Zhao2019, Wang2019atm}. This could lead to doubts, notably when suggestions coming from a DSS deviate significantly from the user's presumption~\cite{sinha2002role}. It has been argued that the lack of explanation regarding the decisions made by ML models and the absence of control over their internal processes are major drawbacks in critical decision-making processes~\cite{Choo2018}. 

The use of visualisation techniques has been shown to provide essential insights and facilitate the understanding of ML models for complex problems~\cite{kulesza2015principles,donoso2018interactive}, particularly towards the users with little ML knowledge~\cite{yosinski2015understanding, kahng2019gan}. Therefore, to address the aforementioned challenges, in this paper, we present a visual support system that allows agricultural experts with little ML knowledge to compare different predictive models by leveraging their domain knowledge. We named this system, AHMoSe for \underline{a}ugmented by \underline{h}uman \underline{mo}del \underline{se}lection. The goals of AHMoSe are: (1) to be able to select models that generalise better to the data and (2) to leverage user understanding of the different models. It can be used with any ML model as it is based on model-agnostic ML interpretation methods, and thus only needs access to the input (i.e., features) and the predicted output of the model. In this paper, we present a case study of AHMoSe in the domain of viticulture.

The contributions of our work in this paper are as follows. First, we present the results of a simulation study indicating how AHMoSe can help in the selection of models that have a better performance than the model that would be selected by an AutoML process. We then present the results of a qualitative evaluation with experts (n=9) from viticulture and ML domains, which indicate that by showing the agreement and disagreement of each model with domain knowledge and enabling users to inspect the models further can promote trust. In addition, the experts identified several potential use cases in agriculture such as understanding how one's products are being characterised by ML models, detecting anomalies in data and unqualified products, and highlighting the problems within the features of a model. The source code of AHMoSe is available at: \url{https://osf.io/6b38w/}.

\section{Related Work}
In this section, we discuss related work in agricultural DSSs, interactive model analysis and human-guided machine learning systems with a focus on systems that allow model comparison and selection.

\subsection{Decision Support Systems in Agriculture}
DSSs have become increasingly popular in the various sub-domains of agriculture~\cite{gutierrez2019review}. Vite.net~\cite{rossi2014addressing}, for instance, supports crop management within vineyards. It provides important information about vine growth, pest control and diseases in grape berries and allows farmers to make informed management decisions. 
AquaCrop~\cite{lorite2013aquadata} supports simulation analysis towards the impact of climate change, especially rainfall, on wheat yield . Similarly, ATLAS~\cite{thierry2017simulating} allows a simulation of crop availability on a landscape across different crop scenarios in relation to pests, diseases and biological control. While these tools can undeniably offer valuable decision supports for agricultural experts, the use of model explanation techniques is limited among agricultural DSSs. As mentioned earlier, black-box systems could often lead to negative perceptions from users when they fail to provide meaningful explanations~\cite{sinha2002role}. We address this by designing a system that can compare and explain the predicted outcomes of various machine learning models in relation to the knowledge of domain experts.

\subsection{Interactive Model Analysis}
Interactive model analysis aims to facilitate the understanding of machine learning models and to help users identify unsatisfactory learning processes through visual analytics. The majority of previous work on interactive model analysis has aimed to support ML experts by allowing them to not only understand and diagnose but also to refine ML models in order to improve their performance and robustness~\cite{LIU201748}. 

\subsubsection{Model Explanations} Research on explainable artificial intelligence (XAI) has recently provided a new range of useful techniques that facilitate the understanding and diagnosis of models. In particular, the visualisation of explanations from model-agnostic interpretation methods, such as LIME~\cite{Ribeiro2016}, SHAP~\cite{Lundberg2017}, DeepVid~\cite{Wang2019} or RuleMatrix~\cite{Ming2019}, allow users to understand better and diagnose different models regardless of the models' internal mechanism. In AHMoSe, we use the SHAP framework to incorporate explanations as its methods show better consistency with human intuition~\cite{Lundberg2017, lundberg2018consistent}. The latter makes it a good fit for our system as the users have to compare these explanations with their domain knowledge. Moreover, the SHAP framework has already been used successfully in the agriculture domain to determine the appropriate time-series length in the prediction of agricultural droughts~\cite{lees2020machine} and to identify the influence of features favourable to the coffee leaf rust disease~\cite{LASSO2020105640}.

\subsubsection{Model Comparison Systems} Although interactive model analysis systems have generally focused on diagnosing or improving a single model, some previous work already has put the spotlight on model comparison and selection. Zhang et al.~\cite{zhang2019manifold} proposed Manifold, a model-agnostic framework aimed at ML experts that provides a visual comparison of ML models that can be used to diagnose and refine ML models on classification and regression tasks. Targeting non-ML experts, as in AHMoSe, B{\"o}gl et al.~\cite{bogl2013visual} proposed TiMoVA, a visual interface to select ARIMA models through interactive visual interfaces based on user stories and iterative domain expert feedback. Unlike AHMoSe, which is model-agnostic, TiMoVa is focused on only a family of models, ARIMA. Besides, M{\"u}hlbacher and Piringer~\cite{muhlbacher2013partition} presented a framework targeting experts from the energy sector that enables visual investigation of patterns to avoid structural assumptions and allows the comparison of different regression models. The system significantly reduced the effort of the experts in building and improving regression models. Unlike the work of M{\"u}hlbacher and Piringer, where the target users have had experience with prediction models, our target users are agricultural experts who have little to no ML experience.

\subsection{Human-Guided Machine Learning}
Some recent work has looked into extending the human-in-the-loop approach to an AutoML context, a research area referred to as human-guided machine learning (HGML), and which tasks analysis and desired system requirements are presented by Gil et al.~\cite{Gil2019}. The end goal of HGML is to develop systems that enable domain experts to use relevant domain knowledge to guide the different steps in an end-to-end AutoML process without the need for a ML expert. The major tasks foreseen by the authors are grouped into three categories: data use, model development, and model interpretation. Here, we focus on four recent systems that fit into the HGML description and have support, besides other tasks, for model comparison and selection, namely TwoRavens~\cite{Honaker2014, Gil2019}, Snowcat~\cite{Cashman2019}, BEAMES~\cite{Das2019}, and Visus~\cite{Santos2019}.

To support model comparison and selection, TwoRavens~\cite{Honaker2014, Gil2019} offers some evaluating metrics (accuracy, mean squared error, or F1 macro) together with a confusion matrix. The same approach is used by Snowcat~\cite{Cashman2019}, with the addition of a residual bar chart for regression models. BEAMES~\cite{Das2019}, that already identifies model comparison as one of the limitations and a possible line of work, allows comparing different models by inspecting their outputs. In contrast to the other three systems, Visus~\cite{Santos2019} supports more detailed explanations. In particular, for classification models, it uses RuleMatrix~\cite{Ming2019} model-agnostic explanations (in addition to a confusion matrix), and for regression models, it uses partial dependence plots~\cite{Friedman2001} to explain every feature, and a confusion scatterplot. Note that neither Visus nor the other three HGML systems use any summary or overview visualisation that eases the comparison process between models explanations, which is one of the design decisions of AHMoSe to improve the comparison process.

Our system, AHMoSe, aims to offer a solution for the comparison of regression models by domain experts with little ML knowledge that uses intuitive model explanations and gives domain-knowledge a primary role on the model comparison and selection. The design of this solution was done so that it could be integrated into any HGML system to improve its support for model understanding, comparison, and selection.

\section{Case Study and Design Process}

To demonstrate AHMoSe and its design process, we use viticulture as an example domain. Many different aspects of the viticulture industry could leverage the use of ML models to generate predictions, e.g., wine quality, grape quality, fermentation duration, and yield. Many of such observations, however, are yearly-based (e.g., yield or grape quality at harvest), which makes it impossible to accelerate the gathering of data. The distribution of the data from year to year can vary a lot due to uncontrollable parameters like weather conditions~\cite{Bendre2015}.

Besides, viticulture researchers often have access to data from just a few vineyards, and grape growers are usually reluctant to share their data. In addition, among all possible metrics that could be measured in viticulture, vineyards often have their own subset of metrics of interest, which can be different from other vineyards. This consequently limits the use of data from different sources together. 

In addition to the limited size of datasets, the black-box nature of ML models has been criticised in the domain~\cite{malherbe2004modeling}. As a result, researchers often rely on knowledge-based decision support systems, such as fuzzy sets~\cite{papageorgiou2016fuzzy}, that use domain knowledge of experts. Similarly, we designed AHMoSe to help with cases where there is a lack of data, and to promote user trust by explaining ML model predictions. AHMoSe combines domain knowledge of the experts with ML models to help domain experts select ML models and to explain the model behaviour as a basis to increase user trust in the model.

Following the nested model for visualisation design proposed by Munzner~\cite{munzner2009nested}, in the following subsections, we (1) characterise the challenges that need to be addressed, (2) analyse the required data transformations and (3) devise tasks for the system.

\subsection{Domain Problem Characterisation}\label{sec:domaincharacterization}
Through several discussion sessions with experts in viticulture, we characterised the different questions that they faced with regards to the previously described limitations:

\textbf{Should I use this model with this data?} Even when the traditional ML metrics (e.g., accuracy, error) are shown and understood by domain experts, it is not clear for them to what data the model would be able to generalise. For example, if a model was trained using data from a nearby vineyard of a different variety, can it be used to make predictions on this one?

\textbf{Why should I trust this ML model?} Users must be provided with an accurate estimation of uncertainties from visual analytics systems so that they can trust the acquired knowledge~\cite{sacha2016role}. Similar to previous work in other domains, domain experts are often very critical when models present outcomes that do not fully match their expectations~\cite{gutierrez2018lada}. In other words, does a model behave according to prior knowledge?  

\textbf{Which model should I select?} When multiple models are available (different models or the same model trained on different data), how can one decide which one is more appropriate to make predictions on their data?  

\subsection{Data Transformation and Abstraction}
\begin{figure}[ht!]
    \centering
    \includegraphics[width=\columnwidth]{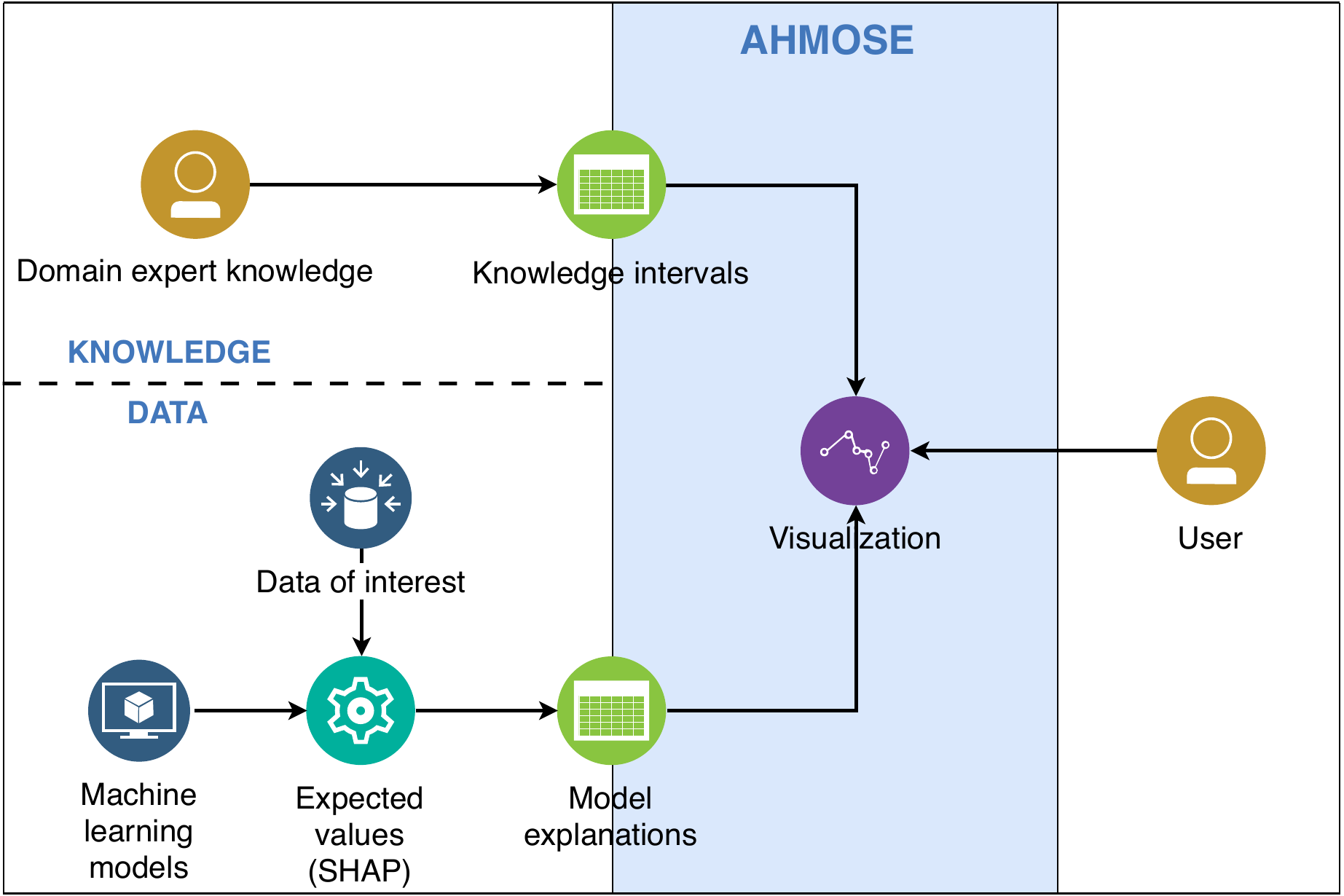}\\
    \caption{An overview of the flow of data in the AHMoSe ecosystem.}
    \label{fig:datapipeline}
\end{figure}

The general transformations and flow of the data are depicted in \autoref{fig:datapipeline}. In this section, we focus on the transformations that should be applied to the raw data to explain models to domain experts. Then, in Section~\ref{sec:ahmoseinterface}, the different transformations and encodings that take place to generate the visualisation, as well as the interactions of the user with the AHMoSe interface, are described. 

\subsubsection{Domain Knowledge}
Previous research has shown the advantages of prompting users to reflect on their prior knowledge~\cite{Kim2017}. AHMoSe expects the domain experts to state their prior knowledge for features explicitly. To this end, they should provide a range where the target feature mean value is expected for as many different intervals of each input feature as they know of.

\subsubsection{Model Explanations}
The reason behind data transformation is to be able to use an explainable ML approach that allows us to know how the output of each model depends on each feature. In particular, we want to know for each observation of our data what effect each feature has on the predicted output. Although a number of algorithms are available for this, we are using the SHAP framework as it shows better consistency with human intuition~\cite{Lundberg2017, lundberg2018consistent}.
 
When obtaining explanations from the SHAP framework for a model and a dataset, one obtains, for each observation of the data, what effect the value of each input feature has on the predicted output (this effect is called the SHAP value). If the model is, for instance, a regression model, this SHAP value indicates how feature increments or decrements affect the output value for that observation. The SHAP framework also provides a base value, which is the average model output over the dataset the model was trained on. This value is used in our system to translate the SHAP value from an effect space to the dimensions of the predicted output, which makes the interpretation and the comparison with domain knowledge easier.

The model explanation data then includes the followings for each model, each feature, and each observation of the data of interest: the value of the feature, the effect of that feature on the model output (SHAP value), and the sum of the SHAP value and the reference value. This last value is very important as it translates the SHAP value from an effect space to the dimensions of the target feature, where we will be able to compare it with the knowledge intervals.  

Most of the tools that visualise model explanations are usually aimed at either classification problems~\cite{Ming2019} or regression problems~\cite{muhlbacher2013partition}. AHMoSe is designed for regression problems where both input and output variables are numeric.

\begin{figure*}[t!]
  \includegraphics[width=\linewidth]{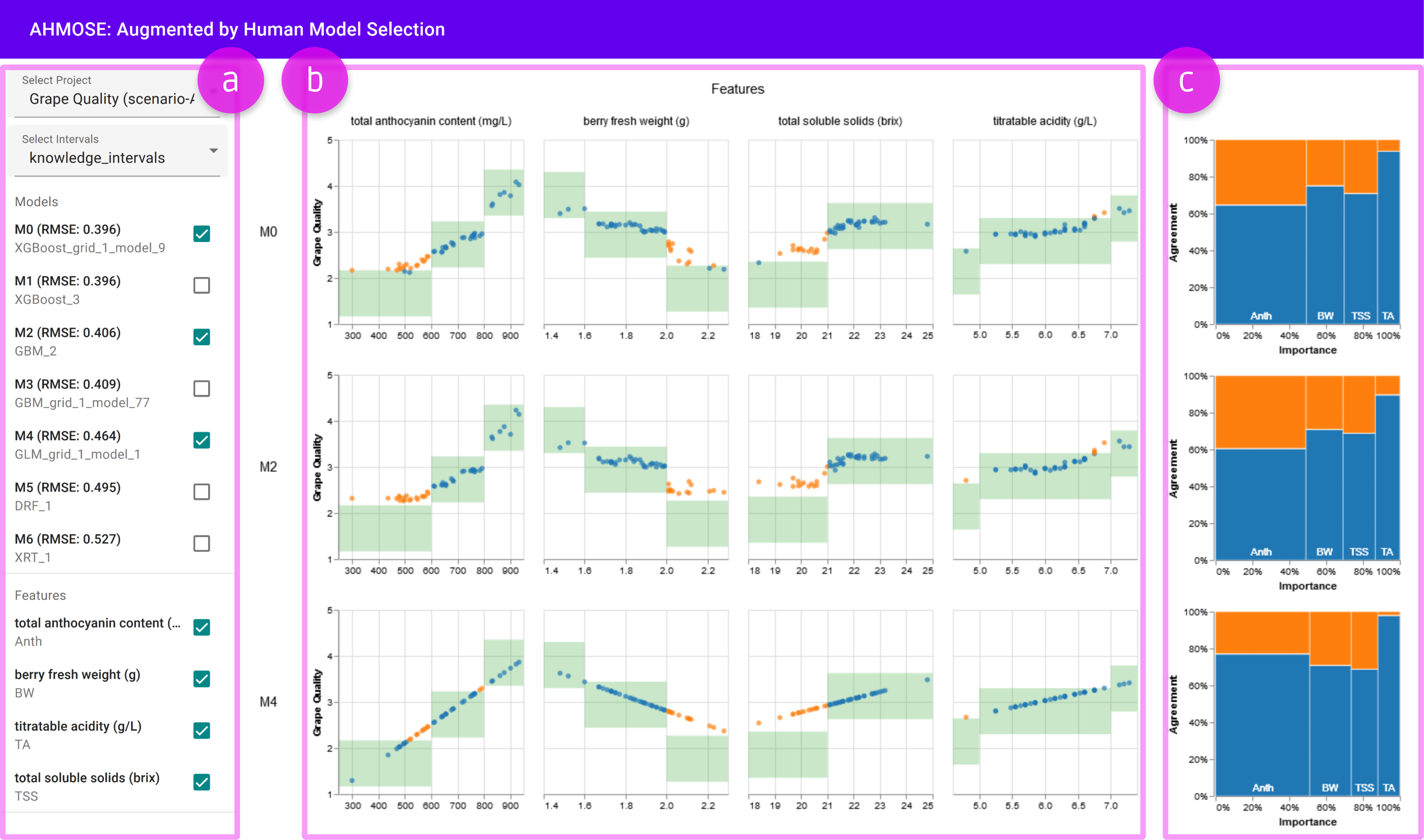}
  \caption{The AHMoSe interface: a) the sidebar controls to select the use case, intervals, models and features to be visualised, b) the scatter plots highlight the comparisons between models' predictions (orange and blue dots) and domain expert knowledge (green rectangles), c) the Marimekko charts indicate the importance of each feature according to the given model, and the agreement between the model and domain expert knowledge.}
  \label{fig:interface}
\end{figure*}

\subsection{Task Abstraction}\label{sec:task}
Based on the different questions devised in Section \ref{sec:domaincharacterization}, using the transformed data (knowledge intervals and model explanations) with AHMoSe should support the following tasks:
\begin{itemize}
    \item[\textbf{T1}.]\textbf{Understand model explanations.} Users should be able to \textit{locate} and \textit{summarise} the explanations of a given model. 
    \item[\textbf{T2}.]\textbf{Identify model bias.} Users should be able to identify if different expected values of model explanations agree with the corresponding knowledge interval, are over-estimations or are under-estimations.
    \item[\textbf{T3}.]\textbf{Compare two different model explanations.} Users should be able to analyse the \textit{similarity} of two different models. 
    \item[\textbf{T4}.]\textbf{Identify a model.} Users should be able to identify a model that has the highest agreement level with knowledge intervals taking into account the importance of each feature.
\end{itemize}

\section{The AHMoSe Interface}\label{sec:ahmoseinterface}

The interface of AHMoSe consists of a sidebar (\autoref{fig:interface}a) and a visualisation panel (\autoref{fig:interface}b and \autoref{fig:interface}c). The visualisation panel consists of a plot matrix that has one row for each of the selected models the user wants to analyse. For each of the rows (i.e., models), two different visualisations, described in the following sections, are shown: a knowledge-agreement dependence plot for each feature (\autoref{fig:interface}b) and a knowledge-agreement summary plot (\autoref{fig:interface}c). The AHMoSe user interface has been developed using React and the visualisations were made with Vega-Lite~\cite{satyanarayan2017vega} and Vega~\cite{satyanarayan2016reactive}.

\subsection{Knowledge-agreement Dependence Plot}
\begin{figure}[ht!]
    \centering
    \includegraphics[width=0.5\columnwidth]{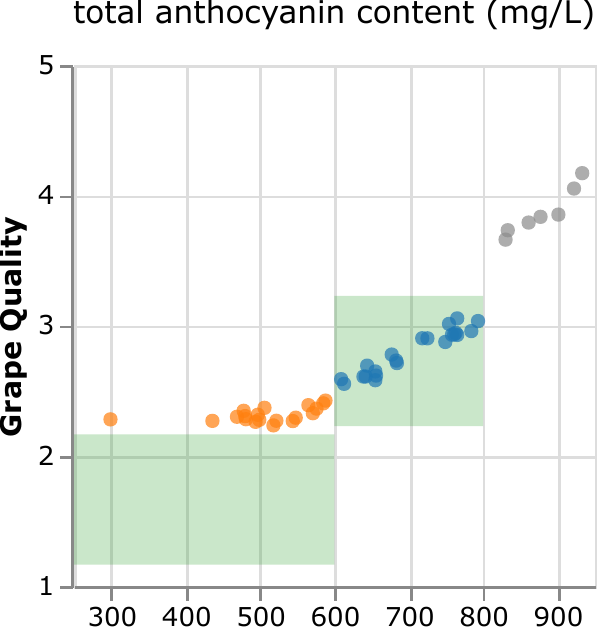}\\
    \caption{Knowledge-agreement dependence plot for a feature (total anthocyanin content).}
    \label{fig:dependanceplots}
\end{figure}

For each model and each feature, a knowledge-agreement dependence plot is shown (e.g., \autoref{fig:dependanceplots}). All these plots are then arranged on a faceted plot matrix, see \autoref{fig:interface}b, where each column corresponds to a feature and each row to a model. This plot has two different layers: a knowledge layer and a model layer. Note that while the model layer is different for each row (model) and column (feature), the knowledge layer is different for each column (feature), but shared by all the models. 

\subsubsection{Knowledge Layer} This layer uses a green rectangle mark for each of the knowledge intervals that correspond to the feature (e.g., total anthocyanin content in \autoref{fig:dependanceplots}). The x-axis limits of each rectangle encode the minimum and maximum of each interval defined by the domain expert. The y-axis limits of each rectangle encode the range where the domain expert expects the target feature mean value to be (e.g., Grape Quality mean value in \autoref{fig:dependanceplots}).

\subsubsection{Model Layer} This layer uses a circle mark for each local explanation corresponding to a specific model and feature (e.g., total anthocyanin content in \autoref{fig:dependanceplots}). The x-position encodes the value of the feature (e.g., total anthocyanin content in \autoref{fig:dependanceplots}) on the observation the local explanation was made on. The y-position encodes the expected value of the target feature (e.g., Grape Quality in \autoref{fig:dependanceplots}) based on the model. Color encodes the agreement (blue) or disagreement (orange) with the corresponding domain expert knowledge. If no knowledge interval covers the value of the feature for a circle, then it is encoded in grey.

The knowledge-agreement dependence plots allow the user to perform task T1 (Understand model explanations) and T2 (Identify model bias). T3 (compare two different model explanations) can also be performed when two models are displayed at a time on the matrix plot. 

\subsection{Knowledge-agreement Summary Plot}
\begin{figure}[ht!]
    \centering
    \includegraphics[width=0.6\columnwidth]{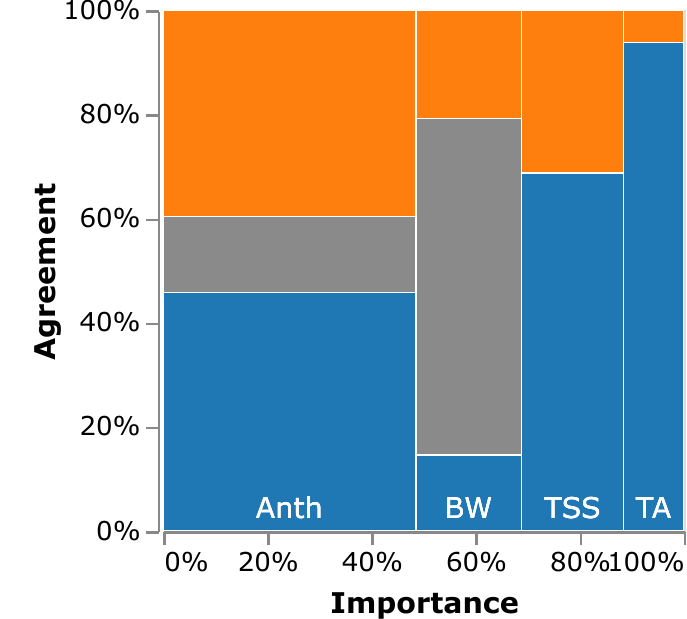}\\
    \caption{Knowledge-agreement summary plot.}
    \label{fig:summaryplot}
\end{figure}

The knowledge-agreement summary plot (e.g., \autoref{fig:summaryplot}) is a Marimekko chart. The width of each group of stacked bars encodes the importance of that feature according to the given model explanations. The height of each stacked bar encodes the percentage of circles of the knowledge-agreement dependence plot of that feature that corresponds to each category: agreement (blue), disagreement (orange), or have no knowledge interval reference (grey). 

The knowledge-agreement summary plot allows the user to perform task T1 (Understand model explanations) and T3 (Compare two different model explanations) as it shows the importance of each feature to the model. Also, the summary plots allow the user to perform T4 (Identify a model).

To perform task T4 (Identify a model), the model with the highest weighted mean agreement (where the weights are the feature importance) should be selected. On the Marimekko chart, this matches with the blue area of the summary plot. Although the comparison of areas is less effective than the comparison of lengths of an alternative bar chart with the weighted mean agreement, the use of Marimekko charts provides transparency to users on how this score is obtained. Also, the importance is still needed for both T1 and T3, so at least two graphs should be added.

\subsection{Sidebar Controls}
The sidebar controls (see \autoref{fig:interface}a) have two different functions: to select which data to use and to filter it to generate the visualisation.

To select which data to load, the user selects a project from a dropdown, which will load the models' explanations and related information and the list of knowledge intervals available. Afterwards, the use of one of the available knowledge intervals for that project can be selected from another dropdown. 

When the data is selected, the sidebar displays a list with the available models and features. The user can select which models and features to visualise. 

\section{Evaluation}\label{sec:usecase}
In this section, we describe the use case of AHMoSe\footnote{\url{https://93degree.github.io/AHMoSe-app/}} to demonstrate the utility of the tool. We use the scenario of grape quality prediction based on~\cite{Tagarakis2014}, where a knowledge-based fuzzy inference system (FIS) is proposed as a model for grape quality in vineyards. Grape quality prediction is representative of many of the common agriculture prediction scenarios, with sparse data as one of the key characteristics.

We first present the results of a simulation study that demonstrates that our approach can enable the selection of a model with better performance than the model proposed by a fully automated AutoML system using a cross-validation approach. We then present the results of a user evaluation with domain experts, which was designed to obtain qualitative feedback on the tool.

\subsection{Simulation Study}\label{sec:simulation}
\subsubsection{Scenario and Data}
In this scenario, the data observations correspond to one grape variety (\textit{Vitis vinifera} cv. Agiorgitiko) on 48 $10 \times 20$ meters cells of the same vineyard situated on Central Greece. The data is available only for the years 2010, 2011, and 2012. The FIS model proposed in~\cite{Tagarakis2014} uses four measures of the grapes taken at harvest for each of the cells as input: total soluble solids (TSS), titratable acidity (TA), total skin anthocyanins (Anth) and berry fresh weight (BW). In order to validate the FIS model, a grape quality score was given by five grape growers and winemakers of the Greek wine industry with experience in Agiorgitiko variety. The grape quality score (GTQ) given by the experts was one of the following values: very poor, poor, average, good, and excellent. Although the GTQ value given by experts is discrete, it is considered as continuous by the FIS model. In the same way, we encoded it from 1 (very poor) to 5 (excellent) for training our regression models and treat it as a continuous value. 

To make the use case scenario realistic, we place ourselves at the harvest time of 2012 for this case study. Thus, we have the four measures on each of the 48 cells available for the three years, but the GTQ grape quality value only of 2010 and 2011. Thus, the models would be trained and validated using only 2010 and 2011 data, and the objective is to use AHMoSe to select the best-suited ML model for predicting the quality of 2012 grapes for each of the 48 cells. The ground truth data for 2012 grape quality will be used in Section~\ref{sec:results} to evaluate the performance improvement obtained when using the model selected with the aid of AHMoSe instead of the model proposed by the fully automated AutoML system. 

\subsubsection{Models}\label{subsec:models}
To construct the different ML models, we used an AutoML system. Using an AutoML system to obtain the models that we are going to use has two advantages:
\begin{itemize}
    \item AutoML systems improve the reproducibility of scientific studies and the fairness of ML model comparisons~\cite{Feurer2019}.
    \item Shows how AHMoSe could be integrated as part of a fully end-to-end human-guided machine learning system.
\end{itemize}

In particular, we have used the AutoML process of the H2O.ai open-source platform through its R API~\cite{landry2019}. The H2O AutoML system trains different models, tuning them using a grid hyperparameter optimisation. In order to evaluate and rank the models, the H2O AutoML process uses by default 5-fold cross-validation for each model in the AutoML to evaluate and rank them. Although H2O.ai supports the use of Deep Neural Networks, we did not use them despite the popularity of this technique~\cite{lecun2015deep}. Our main argument is that the number of parameters in Deep Neural Networks can be easily more than the number of training data points, so it can easily lead to overfitting despite using regularisation techniques such as Frobenius norm or dropout~\cite{goodfellow2016deep}. Furthermore, including the generation of Deep Learning models in the H2O AutoML process leads to results that are not reproducible.

We trained the AutoML function with the 96 observations corresponding to the data from the years 2010 and 2011 and using the default 5-fold cross-validation with the RMSE metric. The cross-validation metrics are used for early stopping and ranking. From the different regression metrics available to rank and evaluate the different results, we use the root mean squared error~(RMSE) throughout this scenario. Although this metric is a bit more challenging to understand for users with little ML knowledge than the mean absolute error~(MAE), the fact that it punishes large errors makes it a good fit. The latter idea is supported by the use of discrete scores by the growers and winemakers.  

In order to reproduce the results, the following parameters (and the default ones for the remaining) should be used on the AutoML function running on an H2O cluster with version 3.26.0.2: 
\begin{itemize}
    \item \texttt{sort\_metric = "RMSE"}
    \item \texttt{stopping\_metric = "RMSE"}
    \item \texttt{seed = 93}
    \item \begin{verbatim}exclude_algos = 
    c("DeepLearning", "StackedEnsemble")
    \end{verbatim}
\end{itemize}
With the previous parameters, the AutoML function generated in approximately 75 seconds 101 different models: 1~DRF, 1~XRT, 1~GLM, 67~GBMs, and 31~XGBoost models. The AutoML process was run on a virtual machine (VM) instance with 4 Intel Xeon Skylake (2.7 GHz) and ~8 GB RAM.

For the case study, only the two best-ranked models of each family based on the RMSE score were selected (for some families, only one is generated by AutoML). In \autoref{tab:models}, the seven selected models are ranked using the AutoML 5-fold cross-validation RMSE score. 

\begin{table}[ht!]
\centering
\caption{The best models of each family (2 if available) with their AutoML 5-fold cross-validation RMSE score and their rank on the AutoML leaderboard.}
\label{tab:models}
\begin{tabular}{llrr}
\toprule
alias & model\_id & RMSE & rank \\ 
\midrule
M0 & XGBoost\_grid\_1\_model\_9 & $0.396$ & 1 \\ 
M1 & XGBoost\_3 & $0.396$ & 2 \\ 
M2 & GBM\_2 & $0.406$ & 4 \\ 
M3 & GBM\_grid\_1\_model\_77 & $0.409$ & 6 \\ 
M4 & GLM\_grid\_1\_model\_1 & $0.464$ & 30 \\ 
M5 & DRF\_1 & $0.495$ & 42 \\ 
M6 & XRT\_1 & $0.527$ & 50 \\ 
\bottomrule
\end{tabular}
\end{table}

\subsubsection{Prior Expert Knowledge} \label{subsec:knowledge}

AHMoSe needs prior expert knowledge on how the target feature (grape quality, GTQ) depends on each input features: total soluble solids (TSS), titratable acidity (TA), total skin anthocyanins (Anth) and berry fresh weight (BW). These prior beliefs should be materialised as regions where the grape quality mean value is expected for different intervals of each feature.  

Although there are different strategies to obtain this prior expert knowledge, to facilitate the reproducibility of the case study and to keep the evaluation not biased to a single expert, we construct the prior expert knowledge from the ``IF-THEN'' rules that are provided by human experts for the FIS construction in~\cite{Tagarakis2014}. 
In \autoref{tab:intervals}, we show how each feature is subdivided in a different number of intervals (between 2 and 4), and each interval is given one of the following labels: L - low, M - medium, H - high or VH - very high. These intervals correspond with the ones defined by the experts in~\cite{Tagarakis2014}. 

\begin{table}[ht!]
\centering
\caption{Knowledge Intervals}
\label{tab:intervals}
\begin{tabular}{@{}llrrr@{}}
\toprule
Feature & Label & Interval       & WQM & GTQ range \\ \midrule
Anth    & L     & {[}200, 600{]} & 1.67 & {[}1.17, 2.17{]} \\
        & M     & (600, 800{]}   & 2.73 & {[}2.23, 3.23{]} \\
        & H     & (800, 1000{]}  & 3.85 & {[}3.35, 4.35{]} \\
        & VH    & (1000, 1400{]} & 4.75 & {[}4.25, 5.00{]} \\ \midrule
BW      & L     & {[}1.0, 1.6{]} & 3.80 & {[}3.30, 4.30{]} \\
        & M     & (1.6, 2.0{]}   & 2.94 & {[}2.44, 3.44{]} \\
        & H     & (2.0, 2.5{]}   & 1.77 & {[}1.27, 2.27{]} \\ \midrule
TSS     & L     & {[}15,21{]}    & 1.86 & {[}1.36, 2.36{]} \\
        & H     & (21,30{]}      & 3.13 & {[}2.63, 3.63{]} \\ \midrule
TA      & L     & {[}3, 5{]}     & 2.14 & {[}1.64, 2.64{]} \\
        & M     & (5, 7{]}       & 2.80 & {[}2.30, 3.30{]} \\
        & H     & (7, 12{]}      & 3.29 & {[}2.79, 3.79{]} \\ \bottomrule
\end{tabular}
\end{table}

To calculate a region where the grape quality mean is expected for each interval (GTQ range column of \autoref{tab:intervals}), we make use of the ``IF-THEN'' rules defined by experts. There are 72 rules, one for each possible combination of the different intervals of each feature. For each combination (input), the rule defines an output (grape quality GTQ). For example, if TSS is high (H), TA is medium (M), Anth is very high (VH), and Berry Weight is low (L), the rule defines a GTQ of 5 (excellent). 

We could obtain the expected mean value of grape quality for each interval of each feature by averaging the output of every rule that includes it. However, such an approach would not lead to a good simulation of a domain expert of our selected vineyard as many of these input combinations never occurred. So to make a more realistic approximation of an expert opinion, we weight each of the rules by the number of times that the rule input combination occurred in the three years we have measures of TSS, TA, Anth, and BW. The W column of \autoref{tab:rules} shows the weight of each rule.  

\begin{table}[ht!]
\centering
\caption{An excerpt of the 72 expert rules. They correspond to the rules with a weight (W) value greater than 0.}\label{tab:rules}
\resizebox{\columnwidth}{!}{%
\begin{tabular}{@{}r|llll|r|lr|llll|r|@{}}
\cmidrule(lr){2-6} \cmidrule(l){9-13}
\textbf{W} & \multicolumn{1}{r}{\textit{TSS}} & \multicolumn{1}{r}{\textit{TA}} & \multicolumn{1}{r}{\textit{Anth}} & \multicolumn{1}{r|}{\textit{BW}} & GTQ &  & \textbf{W} & \textit{TSS} & \textit{TA} & \textit{Anth} & \textit{BW} & \multicolumn{1}{l|}{GTQ} \\ \cmidrule(lr){2-6} \cmidrule(l){9-13} 
\textbf{28} & H & M & M & M & 3 &  & \textbf{2} & L & M & M & L & 3 \\
\textbf{16} & H & M & H & M & 4 &  & \textbf{2} & H & M & VH & L & 5 \\
\textbf{10} & H & M & M & H & 2 &  & \textbf{2} & L & M & L & H & 1 \\
\textbf{10} & L & M & M & M & 2 &  & \textbf{2} & H & L & M & M & 2 \\
\textbf{9} & L & M & L & M & 1 &  & \textbf{2} & H & L & L & M & 1 \\
\textbf{7} & H & H & L & M & 3 &  & \textbf{1} & L & L & M & M & 1 \\
\textbf{6} & H & M & L & H & 1 &  & \textbf{1} & L & L & M & L & 2 \\
\textbf{6} & H & H & M & M & 4 &  & \textbf{1} & L & L & H & L & 3 \\
\textbf{5} & H & M & L & M & 2 &  & \textbf{1} & H & H & VH & M & 5 \\
\textbf{4} & H & H & L & H & 2 &  & \textbf{1} & L & M & H & H & 2 \\
\textbf{4} & H & M & H & L & 5 &  & \textbf{1} & L & M & H & L & 4 \\
\textbf{4} & H & M & VH & M & 5 &  & \textbf{1} & H & L & M & L & 3 \\
\textbf{3} & H & H & H & M & 5 &  & \textbf{1} & H & L & H & H & 2 \\
\textbf{3} & L & H & L & M & 2 &  & \textbf{1} & H & L & VH & H & 3 \\
\textbf{3} & L & M & H & M & 3 &  & \textbf{1} & H & M & H & H & 3 \\
\textbf{3} & H & L & H & M & 3 &  & \textbf{1} & L & L & L & L & 1 \\
\textbf{2} & H & M & M & L & 4 &  &  &  &  &  &  &  \\ \cmidrule(lr){2-6} \cmidrule(l){9-13} 
\end{tabular}%
}
\end{table}

The column weight quality mean (WQM) of \autoref{tab:intervals} shows the obtained weighted mean for each interval of each feature. The prior belief that we consider for each interval is a 0.5 radius interval around this value. So, for example, for medium (M) total anthocyanin content (Anth) values, between 600 and 800 mg/L, the prior belief is that the mean value of grape quality (GTQ) is between 2.23 and 3.23. This can be seen in \autoref{fig:comparison} (corresponding green region).

\subsubsection{Model Understanding and Diagnosis Using AHMoSe}
Once the project and intervals are loaded using the sidebar controls, the user can start exploring the models to understand and diagnose them.

\begin{figure*}[t!]
    \centering
    \includegraphics[width=\textwidth]{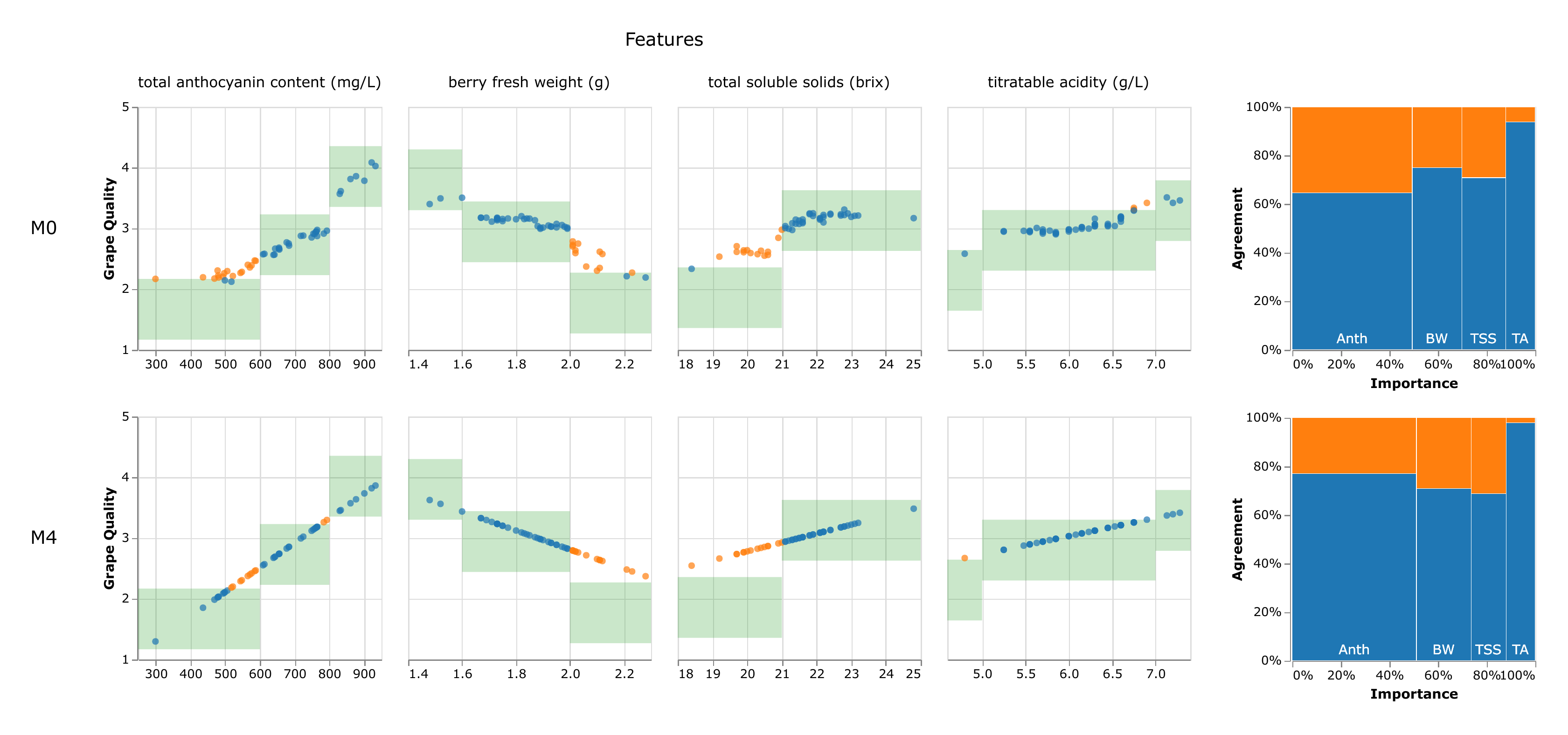}\\
    \caption{Comparison between M0 (XGBoost\_grid\_1\_model\_9) and M4 (GLM\_grid\_1\_model\_1) models.}
    \label{fig:comparison}
\end{figure*}

First, looking at the knowledge-agreement summary plot (2 models can be seen in \autoref{fig:comparison}), a user could observe that Anth (total anthocyanin content) is the most crucial feature for the predictions of every model ($40-50\%$ of the importance). 

Second, there are three intervals where most models expect a higher value than the experts:
\begin{itemize}
    \item Total anthocyanin content (Anth) low interval, from 200 to 600 mg/L. 
    \item Berry fresh weight (BW) high interval, from 2.0 to 2.5~g.
    \item Total soluble solids (TSS) low interval, from 15 to 21 brix.
\end{itemize}
When there is a disagreement on one or more intervals that occurs with many models, it can be a signal of either erroneous expert prior beliefs or a difference in distribution between the training and the data of interest that lead to some models not being able to generalise properly for those values.

\subsubsection{Model Selection Using AHMoSe}\label{sec:results}
Unless other particularities had arisen during the understanding and diagnosis of the models, a sensible approach to select a model using AHMoSe is to identify the one that has a higher weighted mean agreement (i.e., mean of the different agreement levels weighted by the relevance/importance of each feature according to the model). Note that this weighted mean agreement (WMA) is also equivalent to the percentage of the blue area on the summary plot. 

In \autoref{tab:results}, we can see the models ordered by their weighted mean agreement (WMA) value. Also, as we have the real grape quality (GTQ) score given by domain experts in the year 2012, in the column RSME (Test), we have listed their root mean squared error when using the model to make predictions on our test data, the 2012 grape quality. For the sake of completeness, we have also listed the initial AutoML 5-fold cross-validation RMSE score. 

If AHMoSe is not used in this scenario, the selected model would have been the XGBoost\_grid\_1\_model\_9 (M0), as the AutoML system proposes this model due to having the lowest 5-fold cross-validation RMSE of the 101 models that were built. In contrast, using AHMoSe, we can see that the model that has the highest weighted mean agreement (WMA) and thus would be selected is GLM\_grid\_1\_model\_1 (M4). So, by using AHMoSe to support the model selection, we have been able to have a $6.3$ percentage decrease on the root mean squared error (RMSE) for the 2012 test data, from $0.430$ to $0.403$.

\begin{table}
\centering
\caption{Models ordered by their weighted mean agreement (WMA), listed together with their AutoML RMSE 5-fold cross-validation score, and the RMSE score when using the model to make predictions on our test data, the 2012 grape quality.}
\label{tab:results}
\begin{tabular}{llrrr}
\toprule
& & & \multicolumn{2}{c}{RMSE} \\ 
 \cmidrule(lr){4-5}
alias & model & WMA & AutoML & Test \\ 
\midrule
M4 & GLM\_grid\_1\_model\_1 & $0.770$ & $0.464$ & $0.403$ \\ 
M0 & XGBoost\_grid\_1\_model\_9 & $0.714$ & $0.396$ & $0.430$ \\ 
M1 & XGBoost\_3 & $0.693$ & $0.396$ & $0.457$ \\ 
M5 & DRF\_1 & $0.691$ & $0.495$ & $0.567$ \\ 
M6 & XRT\_1 & $0.690$ & $0.527$ & $0.585$ \\ 
M3 & GBM\_grid\_1\_model\_77 & $0.683$ & $0.409$ & $0.526$ \\ 
M2 & GBM\_2 & $0.679$ & $0.406$ & $0.507$ \\ 
\bottomrule
\end{tabular}
\end{table}

\subsubsection{Incomplete Prior Expert Knowledge}
An advantage of AHMoSe over other knowledge-based systems is that the user does not need to know all the features of the models or their interactions. In the description of the scenario of Tagarakis et al.~\cite{Tagarakis2014}, there is some other available data that was not incorporated into their FIS model. Namely, they have measures of the average temperature for the growing season (April 1st to October 31st), and they also have the relative position of each measured cell on the field (the row and column of each one on the grid).

So, we can consider a modification of the previous scenario where we also input these three new features to build and train the models in the AutoML system, but without adding any new knowledge to AHMoSe. From now on, we refer to the previous scenario as A and this one as B.

Following the same rationale as Section~\ref{sec:results} for scenario~B, by using AHMoSe we select GLM\_grid\_1\_model\_1 (B-M4), as it has the highest WMA, instead of the AutoML proposed  XGBoost\_1 (B-M0) model (see models and metrics on \autoref{tab:resultsB}). Thus, in scenario~B, we have been able in this scenario to have a $16.0$ percentage decrease on the root mean squared error (RMSE) for the 2012 test data, from $0.458$ to $0.385$.

As shown here, AHMoSe enables the user to incorporate new features to the prediction model, which can lead, as in this case, to an improvement of the model performance without the need of domain experts to know how they affect the predicted value (grape quality, in our case). Still, the users can compare the features they have knowledge about in the same way, and they can get an understanding of how these new features affect the prediction of their data of interest, based on the explanations shown in AHMoSe.

\begin{table}
\centering
\caption{Scenario~B models ordered by their weighted mean agreement (WMA), listed together with their AutoML RMSE 5-fold cross-validation score, and the RMSE score when using the model to make predictions on our test data, the 2012 grape quality.}
\label{tab:resultsB}
\begin{tabular}{llrrr}
& & & \multicolumn{2}{c}{RMSE} \\ 
 \cmidrule(lr){4-5}
alias & model\textsuperscript{1} & WMA & AutoML & Test \\ 
\midrule
B-M4 & GLM\_grid\_1\_model\_1 & $0.710$ & $0.475$ & $0.385$ \\ 
B-M2 & GBM\_grid\_1\_model\_90 & $0.657$ & $0.425$ & $0.525$ \\ 
B-M1 & XGBoost\_grid\_1\_model\_11 & $0.652$ & $0.412$ & $0.450$ \\ 
B-M0 & XGBoost\_1 & $0.647$ & $0.409$ & $0.458$ \\ 
B-M3 & GBM\_4 & $0.632$ & $0.425$ & $0.520$ \\ 
B-M5 & DRF\_1 & $0.610$ & $0.530$ & $0.589$ \\ 
B-M6 & XRT\_1 & $0.594$ & $0.556$ & $0.624$ \\ 
\bottomrule
\end{tabular}
\vspace{-5mm}
\begin{minipage}{\linewidth}
\textsuperscript{1}\it{Despite some having the same name, the models are different from the ones on scenario~A, as they were trained with additional variables.} \\ 
\end{minipage}
\end{table}

\subsection{Qualitative Feedback} \label{sec:expert_review}

We also conducted an evaluation with experts to gather feedback. We interviewed a total of nine experts, five viticulture experts~(V1 -- V5) and four ML experts~(ML1 -- ML4) who are working across various European research institutes. The experts were first presented with the tool and explained the individual components of the tool. They were then given a few minutes to interact with the tool and to ask questions for further clarifications. If there were no further question, we asked the experts the following 10 open-ended questions. Responses to each of the questions were recorded and later transcribed for analysis.

\begin{enumerate}
  \item How do you imagine the tool being used in real life?
  \item In which situations do you think this tool will be useful?
  \item Which parts of the visualisations make you trust the Machine Learning models?
    \item How did the tool affect your understanding of the influence of features on each model?
  \item How easy was it to use the tool to select a model?
  \item What other word choices (e.g., ``agreement'', ``importance'') would make the visualisation easier to understand?
  \item Which aspects of the visualisations would you change to improve your understanding of the visualisations?
  \item Which aspects of the visualisations would you change to improve your understanding of model explanations?
  \item Which aspects of the visualisations would you change to ease the model selection task?
  \item Do you have suggestions for improvement of the visualisations?
\end{enumerate}

The transcribed data were coded and analysed following the thematic analysis approach~\cite{braun2006using}, which resulted in four main themes: potential use cases, trust, usability, and understandability.

\begin{figure*}[ht!]
    \centering
    \includegraphics[width=\textwidth]{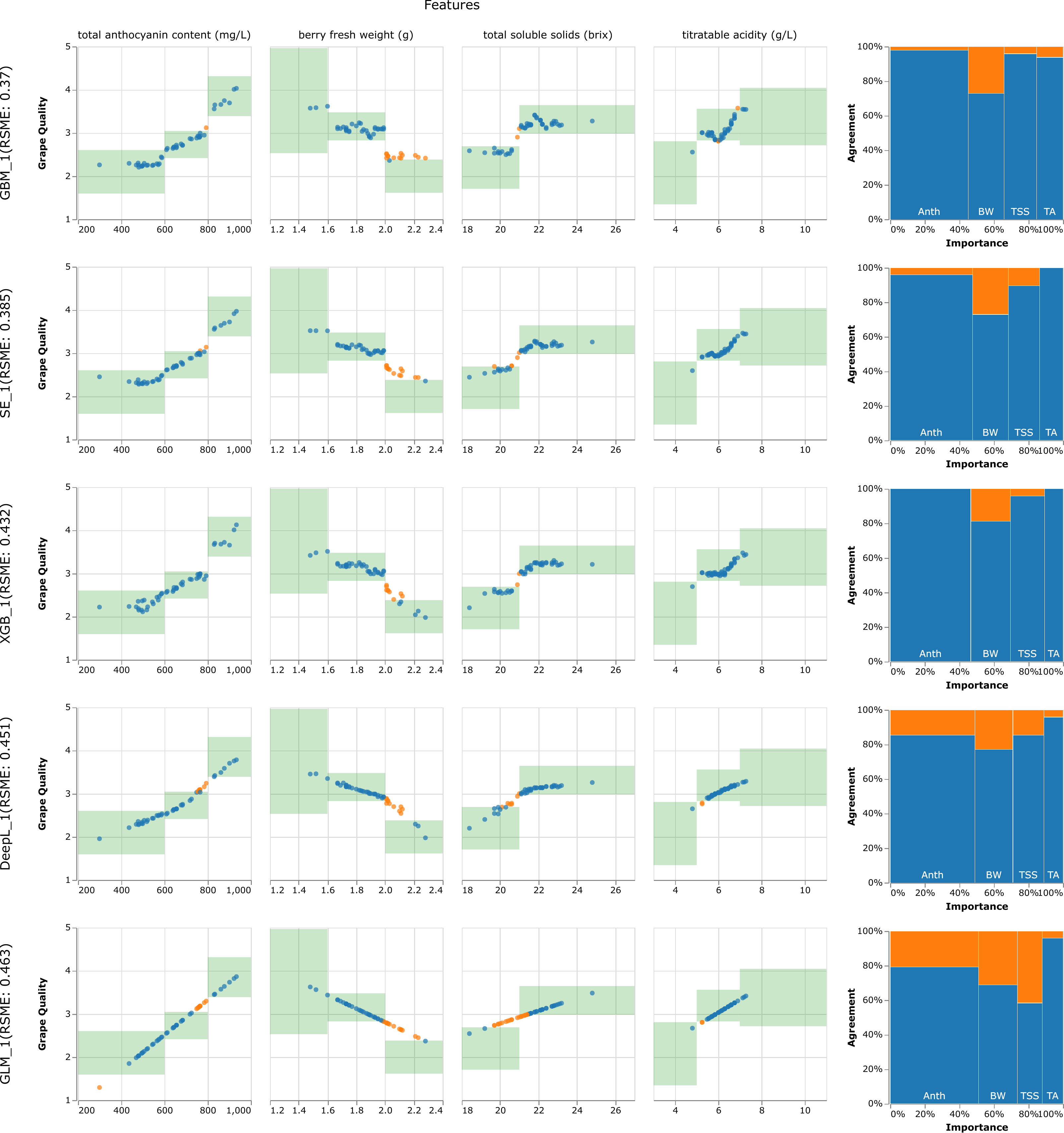}\\
    \caption{Visualisations corresponding to five of the machine learning models available during the evaluation with viticulture and machine learning experts.}
    \label{fig:evaluation}
\end{figure*}

The qualitative evaluation was carried out with different ML models and knowledge intervals than those used for the simulation study previously described in Section~\ref{sec:simulation}. A sample of the specific visualisations shown during the evaluation with viticulture and machine learning experts are shown in \autoref{fig:evaluation}.

\subsubsection{Potential Use Cases} \label{sec:potential_use_cases}
This theme highlights potential use cases that the experts believe can benefit from our tool. All the experts recognised the advantages of using AHMoSe to compare the predictions of various ML models with one's knowledge. For example, one of the viticulture experts explained, ``\textit{If I can put my measurements in there and see how the model characterises my grapes at hand, of my sample, and see for example if my anthocyanin content is 500, how does this relate with the model? Is my quality good or bad?}''~(V1). Another interesting use case mentioned by the viticulture experts was detecting anomalies in data and unqualified products. For example, one expert mentioned, ``\textit{let's say if you have something unexpected happening on the field and you want to see how this correlates with the quality that you are going to get. So you do some measurements, and then you want to see how this is going to translate.}''~(V2).

The ML experts, unsurprisingly, were more focused on the technical process behind each model. While they recognised the advantages of the tool for viticulture experts, they specifically pointed out the case which requires a viticulture expert to select a suitable model and better understand the features. For example, one expert explained, ``\textit{Of course, there are models that are clearly unsuitable but they show where the problems are, on which feature we have a problem in the ML algorithm. I don't know how to exploit this information but it is clear that the berry fresh weight is a difficult feature to use in all models.}''~(ML2). This expert also pointed out the utility of this tool in anomaly detection such as, ``\textit{A domain expert could ask, `I didn't know that this feature, berry fresh weight, when between 2 and 2.4, the grape quality is difficult to predict. What is happening in that range?' In other words...anomaly detection.}''~(ML2). In~\autoref{fig:evaluation}, one can see the described discrepancy between five of the models and the knowledge intervals for the berry fresh weight feature when its values are between 2 and 2.4, as highlighted by the orange coloured circles in the second column of the plot matrix. The position of many circles above the knowledge interval indicates that, compared to the expert knowledge, the models overestimate the quality of grapes with these berry fresh weight values. Thus, it appears that anomaly detection is a potential use case the experts envision the tool being used the most, in addition to quality prediction. Finally, the experts also suggested to give the user more control, such as configuring different models for each threshold~(ML4). 

\subsubsection{Trust}
The trust of users in a model often plays an important role for it to be adopted. Explaining a model's predictions, in a way that it is easily understandable, is perhaps the most important aspect for earning a user's trust~\cite{Ribeiro2016}. 
This theme presents the aspects of our visualisations that may help users trust in the given ML models. All of the nine participants mentioned that the ability to see dis/agreements between models' predictions and an expert's knowledge can help them inspect further and thus promote trust. A viticulture expert explained, ``\textit{The thing that makes us trust the models is the fact that most of the time, there is a good agreement between the values predicted by the model and the ones obtained for the knowledge of the experts.}''~(V5). Another viticulture expert described, ``\textit{You actually help the model with some knowledge from the experts, so when I see for example that my basic knowledge agrees with the model, then we know that the model has some degree of certainty.}''~(V1). However, the same expert also emphasised that we ``\textit{cannot always take for granted and trust it with our eyes closed.}''~(V1). The ML experts also had a similar remark regarding trust. For example, one explained that ``\emph{ [...] this visualisation is telling me an insight on what to look at. They are not going to take any decision or try to explain me why there is this problem; just telling me there could be something going on here}''~(ML1). It appears that although the experts are aware of uncertainties, they recognise the potentials of visualisations and the ability to improve predictions with the knowledge of domain experts.

\subsubsection{Usability}
The usability theme focuses on the perceived ability of the tool to be used in real-life. As previously mentioned in Section \ref{sec:potential_use_cases}, all the experts saw usability of the tool in various scenarios. However, they also expressed certain concerns and provided suggestions for improvement. For example, one viticulture expert mentioned ``\textit{it definitely needed a person there like [researcher name] to explain it a bit, the logic behind the models, what the green areas depict, the importance in the agreement of the model. It wasn't difficult to use it but definitely someone should be there to explain exactly what it is. But it wasn't difficult after that.}''~(V1). Adding more textual explanations and legends may help mitigate this limitation. 

The ML experts also provided some interesting suggestions to improve the usability of the tool. These include sticky labels and using a mouse-over function to explain the models briefly and to show various metrics of the models. For example, one expert explained, ``\textit{When you slide over [hover] a model, give more details because I can probably imagine what the acronyms stand for but they are basically capital letters and numbers and also add some explanations about the metrics you use. Is RMSE 0.37 high or low, with respect to which reference value we are looking at?}''~(ML1). A second ML expert suggested, ``\textit{What I would have on a specific plot, for example the berry fresh weight, is what the size of the error is of the orange dots. I mean the percentage of the total error you get, in terms of performance. What is the error that they bring to the total quality? Maybe you should give some insight of how big it is, how important it is to fix this behaviour for specific classes of visualisation.}''~(ML2). A third ML expert highlighted the importance of also showing the source of the data, whether is it model or human-generated. As these suggestions have highlighted, it appears that usability and understanding of the components may be correlated. In Section \ref{sec:understandability}, we further present the understandability of the various components in our tool.

\subsubsection{Understandability} \label{sec:understandability}
This theme focuses on how easy it was to understand the various components of the tool and how we can improve them. While all the experts reported that they understood the intentions of the tool, they also mentioned that the Marimekko charts (Figure \ref{fig:interface}c) could be improved. For example, one viticulture expert explained that he wanted to see whether ``\textit{[...] this model is in the agreement with the basic bibliography. Is it in the agreement with the expert and his/her knowledge?}''~(V1). One suggestion to show this agreement was to introduce confidence intervals~(ML4). Two of the viticulture experts suggested that explicitly showing the success rate of each model may be helpful. For example, ``\textit{The right part may be a little bit confusing. And it would be useful to have a final number [of] the percentage of success.}''~(V2). The tool currently shows the Root-Mean-Square Error (RMSE) for each model, which may be confusing for the experts who do not have an ML background.

One ML expert stated that the Marimekko charts could be better explained by showing the importance and agreement in relation to a particular baseline, for example, ``\textit{importance of a feature, and agreement with human assessor or some description about that.}''~(ML1). Another ML expert suggested that showing more statistics or even the underlying model (although this could get too technical) may help understand the models better. For example, ``\textit{[...] what is the fraction of the error per feature. Given the wrong answer on each feature, the fraction of the error computed on that specific dimension but maybe this is too technical.}''~(ML2). Surely, understandability of the models, visualisations and features may differ between individual experts depending on their background. However, we believe that the suggestions provided by the ML experts could also help improve understandability for viticulture experts.

\section{Conclusions and Future Work}

In this article, we presented AHMoSe, a visual support system for model selection, in particular, regression models. Our goal was to help users integrate their knowledge for model selection in an intuitive way since several tools perform automatic model selection, but in certain domains, the users need further inspection of results before turning prediction results into decision making. We provided details of the design process that justify important implementation decisions of AHMoSe. Results of a simulation study showed that combining expert knowledge with ML models enables to select a model with better performance than the model that would be selected by an AutoML system. Results of our qualitative evaluation indicated that seeing dis/agreements and inspecting further into the models can promote trust of domain experts. Experts recognised risks and uncertainties within predictions, but also the potentials of the visualisation. Besides, the experts identified several potential use cases such as understanding how the product (e.g., grapes) may be characterised by different models, detecting anomalies in data and unqualified products, and highlighting the problems within the features of a model. One limitation of our tool is that it currently only deals with predicting continuous variables (regression). In future work, we will extend it to deal with classification problems. Moreover, we will also investigate other ways to incorporate user feedback beyond the knowledge intervals. One possibility in this line of research is using weak supervision~\cite{Ratner2019}, in order to make the most out of sometimes scarce but rich expert feedback.

\section*{Acknowledgement}
Funding for this research has been provided by the \emph{European Union's Horizon 2020} research and innovation programme: BigDataGrapes Project (grant number 780751) and the \emph{FWO - Research Foundation Flanders} (grant number G0A3319N). Author Denis Parra was funded by \emph{ANID - Millennium Science Initiative Program} (code ICN17$\_$002) and by \emph{ANID, FONDECYT} (grant number 1191791).

\bibliography{mybibfile}

\begin{thebibliography}{10}
\expandafter\ifx\csname url\endcsname\relax
  \def\url#1{\texttt{#1}}\fi
\expandafter\ifx\csname urlprefix\endcsname\relax\def\urlprefix{URL }\fi
\expandafter\ifx\csname href\endcsname\relax
  \def\href#1#2{#2} \def\path#1{#1}\fi

\bibitem{gutierrez2019review}
F.~Gutiérrez, N.~N. Htun, F.~Schlenz, A.~Kasimati, K.~Verbert,
  \href{https://doi.org/10.1016/j.compag.2019.05.053}{A review of
  visualisations in agricultural decision support systems: {An} {HCI}
  perspective}, Computers and Electronics in Agriculture 163 (2019) 104844.
\newblock \href {http://dx.doi.org/10.1016/j.compag.2019.05.053}
  {\path{doi:10.1016/j.compag.2019.05.053}}.
\newline\urlprefix\url{https://doi.org/10.1016/j.compag.2019.05.053}

\bibitem{Kaul2017}
A.~Kaul, S.~Maheshwary, V.~Pudi,
  \href{https://doi.org/10.1109/icdm.2017.31}{{AutoLearn} --- automated feature
  generation and selection}, in: 2017 IEEE International Conference on Data
  Mining (ICDM), IEEE, 2017, pp. 217--226.
\newblock \href {http://dx.doi.org/10.1109/icdm.2017.31}
  {\path{doi:10.1109/icdm.2017.31}}.
\newline\urlprefix\url{https://doi.org/10.1109/icdm.2017.31}

\bibitem{Katz2016}
G.~Katz, E.~C.~R. Shin, D.~Song,
  \href{https://doi.org/10.1109/icdm.2016.0123}{{ExploreKit:} {Automatic}
  feature generation and selection}, in: 2016 IEEE 16th International
  Conference on Data Mining (ICDM), IEEE, 2016, pp. 979--984.
\newblock \href {http://dx.doi.org/10.1109/icdm.2016.0123}
  {\path{doi:10.1109/icdm.2016.0123}}.
\newline\urlprefix\url{https://doi.org/10.1109/icdm.2016.0123}

\bibitem{Feurer2019}
M.~Feurer, F.~Hutter,
  \href{https://doi.org/10.1007/978-3-030-05318-5_1}{Hyperparameter
  optimization}, in: F.~Hutter, L.~Kotthoff, J.~Vanschoren (Eds.), Automated
  Machine Learning, Springer, Cham, 2019, pp. 3--33.
\newblock \href {http://dx.doi.org/10.1007/978-3-030-05318-5_1}
  {\path{doi:10.1007/978-3-030-05318-5_1}}.
\newline\urlprefix\url{https://doi.org/10.1007/978-3-030-05318-5_1}

\bibitem{Elsken2019}
T.~Elsken, J.~H. Metzen, F.~Hutter,
  \href{https://doi.org/10.1007/978-3-030-05318-5_3}{Neural architecture
  search}, in: F.~Hutter, L.~Kotthoff, J.~Vanschoren (Eds.), Automated Machine
  Learning, Springer, Cham, 2019, pp. 63--77.
\newblock \href {http://dx.doi.org/10.1007/978-3-030-05318-5_3}
  {\path{doi:10.1007/978-3-030-05318-5_3}}.
\newline\urlprefix\url{https://doi.org/10.1007/978-3-030-05318-5_3}

\bibitem{hall2019}
P.~Hall, M.~Kurka, A.~Bartz,
  \href{http://docs.h2o.ai/driverless-ai/latest-stable/docs/booklets/DriverlessAIBooklet.pdf}{Using
  {H2O} Driverless {AI.} Version 1.9.2.1.}, H2O.ai, Inc., 2021.
\newline\urlprefix\url{http://docs.h2o.ai/driverless-ai/latest-stable/docs/booklets/DriverlessAIBooklet.pdf}

\bibitem{domingos1999role}
P.~Domingos, \href{https://doi.org/10.1023/a:1009868929893}{The role of
  {Occam's} razor in knowledge discovery}, Data Mining and Knowledge Discovery
  3~(4) (1999) 409--425.
\newblock \href {http://dx.doi.org/10.1023/a:1009868929893}
  {\path{doi:10.1023/a:1009868929893}}.
\newline\urlprefix\url{https://doi.org/10.1023/a:1009868929893}

\bibitem{Gil2019}
Y.~Gil, J.~Honaker, S.~Gupta, Y.~Ma, V.~D'Orazio, D.~Garijo, S.~Gadewar,
  Q.~Yang, N.~Jahanshad, \href{https://doi.org/10.1145/3301275.3302324}{Towards
  human-guided machine learning}, in: Proceedings of the 24th International
  Conference on Intelligent User Interfaces, IUI '19, ACM, New York, NY, USA,
  2019, pp. 614--624.
\newblock \href {http://dx.doi.org/10.1145/3301275.3302324}
  {\path{doi:10.1145/3301275.3302324}}.
\newline\urlprefix\url{https://doi.org/10.1145/3301275.3302324}

\bibitem{Papadopoulus2011}
A.~Papadopoulos, D.~Kalivas, T.~Hatzichristos,
  \href{https://doi.org/10.1016/j.compag.2011.06.007}{Decision support system
  for nitrogen fertilization using fuzzy theory}, Computers and Electronics in
  Agriculture 78~(2) (2011) 130--139.
\newblock \href {http://dx.doi.org/10.1016/j.compag.2011.06.007}
  {\path{doi:10.1016/j.compag.2011.06.007}}.
\newline\urlprefix\url{https://doi.org/10.1016/j.compag.2011.06.007}

\bibitem{accorsi2014hydroqual}
P.~Accorsi, N.~Lalande, M.~Fabregue, A.~Braud, P.~Poncelet, A.~Sallaberry,
  S.~Bringay, M.~Teisseire, F.~Cernesson, F.~Le~Ber,
  \href{https://doi.org/10.1109/vast.2014.7042488}{{HydroQual:} {Visual}
  analysis of river water quality}, in: 2014 IEEE Conference on Visual
  Analytics Science and Technology (VAST), IEEE, 2014, pp. 123--132.
\newblock \href {http://dx.doi.org/10.1109/vast.2014.7042488}
  {\path{doi:10.1109/vast.2014.7042488}}.
\newline\urlprefix\url{https://doi.org/10.1109/vast.2014.7042488}

\bibitem{han2017climate}
E.~Han, A.~V. Ines, W.~E. Baethgen,
  \href{https://doi.org/10.1016/j.envsoft.2017.06.024}{Climate-agriculture-modeling
  and decision tool {(CAMDT):} {A} software framework for climate risk
  management in agriculture}, Environmental Modelling \& Software 95 (2017)
  102--114.
\newblock \href {http://dx.doi.org/10.1016/j.envsoft.2017.06.024}
  {\path{doi:10.1016/j.envsoft.2017.06.024}}.
\newline\urlprefix\url{https://doi.org/10.1016/j.envsoft.2017.06.024}

\bibitem{rossi2014addressing}
V.~Rossi, F.~Salinari, S.~Poni, T.~Caffi, T.~Bettati,
  \href{https://doi.org/10.1016/j.compag.2013.10.011}{Addressing the
  implementation problem in agricultural decision support systems: {The}
  example of vite.net\textregistered}, Computers and Electronics in Agriculture
  100 (2014) 88--99.
\newblock \href {http://dx.doi.org/10.1016/j.compag.2013.10.011}
  {\path{doi:10.1016/j.compag.2013.10.011}}.
\newline\urlprefix\url{https://doi.org/10.1016/j.compag.2013.10.011}

\bibitem{tayyebi2016smartscape}
A.~Tayyebi, T.~D. Meehan, J.~Dischler, G.~Radloff, M.~Ferris, C.~Gratton,
  \href{https://doi.org/10.1016/j.compag.2015.12.003}{{SmartScape$^{TM}$:} {A}
  web-based decision support system for assessing the tradeoffs among multiple
  ecosystem services under crop-change scenarios}, Computers and Electronics in
  Agriculture 121 (2016) 108--121.
\newblock \href {http://dx.doi.org/10.1016/j.compag.2015.12.003}
  {\path{doi:10.1016/j.compag.2015.12.003}}.
\newline\urlprefix\url{https://doi.org/10.1016/j.compag.2015.12.003}

\bibitem{Zhao2019}
R.~Zhao, R.~Yan, Z.~Chen, K.~Mao, P.~Wang, R.~X. Gao,
  \href{https://doi.org/10.1016/j.ymssp.2018.05.050}{Deep learning and its
  applications to machine health monitoring}, Mechanical Systems and Signal
  Processing 115 (2019) 213--237.
\newblock \href {http://dx.doi.org/10.1016/j.ymssp.2018.05.050}
  {\path{doi:10.1016/j.ymssp.2018.05.050}}.
\newline\urlprefix\url{https://doi.org/10.1016/j.ymssp.2018.05.050}

\bibitem{Wang2019atm}
Q.~Wang, Y.~Ming, Z.~Jin, Q.~Shen, D.~Liu, M.~J. Smith, K.~Veeramachaneni,
  H.~Qu, \href{https://doi.org/10.1145/3290605.3300911}{{ATMSeer:} {Increasing}
  transparency and controllability in automated machine learning}, in:
  Proceedings of the 2019 CHI Conference on Human Factors in Computing Systems,
  CHI '19, ACM, New York, NY, USA, 2019, pp. 681:1--681:12.
\newblock \href {http://dx.doi.org/10.1145/3290605.3300911}
  {\path{doi:10.1145/3290605.3300911}}.
\newline\urlprefix\url{https://doi.org/10.1145/3290605.3300911}

\bibitem{sinha2002role}
R.~Sinha, K.~Swearingen, \href{https://doi.org/10.1145/506443.506619}{The role
  of transparency in recommender systems}, in: CHI '02 extended abstracts on
  Human factors in computing systems - CHI '02, ACM, 2002, pp. 830--831.
\newblock \href {http://dx.doi.org/10.1145/506443.506619}
  {\path{doi:10.1145/506443.506619}}.
\newline\urlprefix\url{https://doi.org/10.1145/506443.506619}

\bibitem{Choo2018}
J.~Choo, S.~Liu, \href{https://doi.org/10.1109/mcg.2018.042731661}{Visual
  analytics for explainable deep learning}, IEEE Computer Graphics and
  Applications 38~(4) (2018) 84--92.
\newblock \href {http://dx.doi.org/10.1109/mcg.2018.042731661}
  {\path{doi:10.1109/mcg.2018.042731661}}.
\newline\urlprefix\url{https://doi.org/10.1109/mcg.2018.042731661}

\bibitem{kulesza2015principles}
T.~Kulesza, M.~Burnett, W.-K. Wong, S.~Stumpf,
  \href{https://doi.org/10.1145/2678025.2701399}{Principles of explanatory
  debugging to personalize interactive machine learning}, in: Proceedings of
  the 20th International Conference on Intelligent User Interfaces, ACM, 2015,
  pp. 126--137.
\newblock \href {http://dx.doi.org/10.1145/2678025.2701399}
  {\path{doi:10.1145/2678025.2701399}}.
\newline\urlprefix\url{https://doi.org/10.1145/2678025.2701399}

\bibitem{donoso2018interactive}
I.~Donoso-Guzmán, D.~Parra, \href{https://doi.org/10.1145/3172944.3172953}{An
  interactive relevance feedback interface for evidence-based health care}, in:
  23rd International Conference on Intelligent User Interfaces, ACM, 2018, pp.
  103--114.
\newblock \href {http://dx.doi.org/10.1145/3172944.3172953}
  {\path{doi:10.1145/3172944.3172953}}.
\newline\urlprefix\url{https://doi.org/10.1145/3172944.3172953}

\bibitem{yosinski2015understanding}
J.~Yosinski, J.~Clune, A.~Nguyen, T.~Fuchs, H.~Lipson, Understanding neural
  networks through deep visualization (2015).
\newblock \href {http://arxiv.org/abs/1506.06579} {\path{arXiv:1506.06579}}.

\bibitem{kahng2019gan}
M.~Kahng, N.~Thorat, D.~H.~P. Chau, F.~B. Viegas, M.~Wattenberg,
  \href{https://doi.org/10.1109/tvcg.2018.2864500}{{GAN} lab: {Understanding}
  complex deep generative models using interactive visual experimentation},
  IEEE Transactions on Visualization and Computer Graphics 25~(1) (2019)
  310--320.
\newblock \href {http://dx.doi.org/10.1109/tvcg.2018.2864500}
  {\path{doi:10.1109/tvcg.2018.2864500}}.
\newline\urlprefix\url{https://doi.org/10.1109/tvcg.2018.2864500}

\bibitem{lorite2013aquadata}
I.~Lorite, M.~García-Vila, C.~Santos, M.~Ruiz-Ramos, E.~Fereres,
  \href{https://doi.org/10.1016/j.compag.2013.05.010}{{AquaData} and {AquaGIS:}
  {Two} computer utilities for temporal and spatial simulations of
  water-limited yield with {AquaCrop}}, Computers and Electronics in
  Agriculture 96 (2013) 227--237.
\newblock \href {http://dx.doi.org/10.1016/j.compag.2013.05.010}
  {\path{doi:10.1016/j.compag.2013.05.010}}.
\newline\urlprefix\url{https://doi.org/10.1016/j.compag.2013.05.010}

\bibitem{thierry2017simulating}
H.~Thierry, A.~Vialatte, J.-P. Choisis, B.~Gaudou, H.~Parry, C.~Monteil,
  \href{https://doi.org/10.1016/j.ecoinf.2017.05.006}{Simulating
  spatially-explicit crop dynamics of agricultural landscapes: {The} {ATLAS}
  simulator}, Ecological Informatics 40 (2017) 62--80.
\newblock \href {http://dx.doi.org/10.1016/j.ecoinf.2017.05.006}
  {\path{doi:10.1016/j.ecoinf.2017.05.006}}.
\newline\urlprefix\url{https://doi.org/10.1016/j.ecoinf.2017.05.006}

\bibitem{LIU201748}
S.~Liu, X.~Wang, M.~Liu, J.~Zhu,
  \href{https://doi.org/10.1016/j.visinf.2017.01.006}{Towards better analysis
  of machine learning models: {A} visual analytics perspective}, Visual
  Informatics 1~(1) (2017) 48--56.
\newblock \href {http://dx.doi.org/10.1016/j.visinf.2017.01.006}
  {\path{doi:10.1016/j.visinf.2017.01.006}}.
\newline\urlprefix\url{https://doi.org/10.1016/j.visinf.2017.01.006}

\bibitem{Ribeiro2016}
M.~T. Ribeiro, S.~Singh, C.~Guestrin,
  \href{https://doi.org/10.1145/2939672.2939778}{{"Why} should i trust you?"},
  in: Proceedings of the 22nd ACM SIGKDD International Conference on Knowledge
  Discovery and Data Mining, KDD '16, ACM, New York, NY, USA, 2016, pp.
  1135--1144.
\newblock \href {http://dx.doi.org/10.1145/2939672.2939778}
  {\path{doi:10.1145/2939672.2939778}}.
\newline\urlprefix\url{https://doi.org/10.1145/2939672.2939778}

\bibitem{Lundberg2017}
S.~M. Lundberg, S.~Lee,
  \href{https://proceedings.neurips.cc/paper/2017/hash/8a20a8621978632d76c43dfd28b67767-Abstract.html}{A
  unified approach to interpreting model predictions}, in: I.~Guyon, U.~von
  Luxburg, S.~Bengio, H.~M. Wallach, R.~Fergus, S.~Vishwanathan, R.~Garnett
  (Eds.), Advances in Neural Information Processing Systems 30: Annual
  Conference on Neural Information Processing Systems 2017, December 4-9, 2017,
  Long Beach, CA, {USA}, 2017, pp. 4765--4774.
\newline\urlprefix\url{https://proceedings.neurips.cc/paper/2017/hash/8a20a8621978632d76c43dfd28b67767-Abstract.html}

\bibitem{Wang2019}
J.~Wang, L.~Gou, W.~Zhang, H.~Yang, H.-W. Shen,
  \href{https://doi.org/10.1109/tvcg.2019.2903943}{{DeepVID:} {Deep} visual
  interpretation and diagnosis for image classifiers via knowledge
  distillation}, IEEE Transactions on Visualization and Computer Graphics
  25~(6) (2019) 2168--2180.
\newblock \href {http://dx.doi.org/10.1109/tvcg.2019.2903943}
  {\path{doi:10.1109/tvcg.2019.2903943}}.
\newline\urlprefix\url{https://doi.org/10.1109/tvcg.2019.2903943}

\bibitem{Ming2019}
Y.~Ming, H.~Qu, E.~Bertini,
  \href{https://doi.org/10.1109/tvcg.2018.2864812}{{RuleMatrix:} {Visualizing}
  and understanding classifiers with rules}, IEEE Transactions on Visualization
  and Computer Graphics 25~(1) (2019) 342--352.
\newblock \href {http://dx.doi.org/10.1109/tvcg.2018.2864812}
  {\path{doi:10.1109/tvcg.2018.2864812}}.
\newline\urlprefix\url{https://doi.org/10.1109/tvcg.2018.2864812}

\bibitem{lundberg2018consistent}
S.~M. Lundberg, G.~G. Erion, S.-I. Lee, Consistent individualized feature
  attribution for tree ensembles (2019).
\newblock \href {http://arxiv.org/abs/1802.03888} {\path{arXiv:1802.03888}}.

\bibitem{lees2020machine}
T.~Lees, G.~Tseng, O.~Solar, A.~Hernandez-Garcia, C.~Atzberger, S.~Dadson,
  S.~Reece, A machine learning pipeline to predict vegetation health, in:
  Eighth International Conference on Learning Representations, 2020, pp. 1--5.

\bibitem{LASSO2020105640}
E.~Lasso, D.~C. Corrales, J.~Avelino, E.~de~Melo Virginio~Filho, J.~C.
  Corrales, \href{https://doi.org/10.1016/j.compag.2020.105640}{Discovering
  weather periods and crop properties favorable for coffee rust incidence from
  feature selection approaches}, Computers and Electronics in Agriculture 176
  (2020) 105640.
\newblock \href {http://dx.doi.org/10.1016/j.compag.2020.105640}
  {\path{doi:10.1016/j.compag.2020.105640}}.
\newline\urlprefix\url{https://doi.org/10.1016/j.compag.2020.105640}

\bibitem{zhang2019manifold}
J.~Zhang, Y.~Wang, P.~Molino, L.~Li, D.~S. Ebert,
  \href{https://doi.org/10.1109/tvcg.2018.2864499}{Manifold: {A} model-agnostic
  framework for interpretation and diagnosis of machine learning models}, IEEE
  Transactions on Visualization and Computer Graphics 25~(1) (2019) 364--373.
\newblock \href {http://dx.doi.org/10.1109/tvcg.2018.2864499}
  {\path{doi:10.1109/tvcg.2018.2864499}}.
\newline\urlprefix\url{https://doi.org/10.1109/tvcg.2018.2864499}

\bibitem{bogl2013visual}
M.~Bogl, W.~Aigner, P.~Filzmoser, T.~Lammarsch, S.~Miksch, A.~Rind,
  \href{https://doi.org/10.1109/tvcg.2013.222}{Visual analytics for model
  selection in time series analysis}, IEEE Transactions on Visualization and
  Computer Graphics 19~(12) (2013) 2237--2246.
\newblock \href {http://dx.doi.org/10.1109/tvcg.2013.222}
  {\path{doi:10.1109/tvcg.2013.222}}.
\newline\urlprefix\url{https://doi.org/10.1109/tvcg.2013.222}

\bibitem{muhlbacher2013partition}
T.~Muhlbacher, H.~Piringer, \href{https://doi.org/10.1109/tvcg.2013.125}{A
  partition-based framework for building and validating regression models},
  IEEE Transactions on Visualization and Computer Graphics 19~(12) (2013)
  1962--1971.
\newblock \href {http://dx.doi.org/10.1109/tvcg.2013.125}
  {\path{doi:10.1109/tvcg.2013.125}}.
\newline\urlprefix\url{https://doi.org/10.1109/tvcg.2013.125}

\bibitem{Honaker2014}
J.~Honaker, V.~D'Orazio,
  \href{http://ceur-ws.org/Vol-1210/datawiz2014\_05.pdf}{Statistical modeling
  by gesture: {A} graphical, browser-based statistical interface for data
  repositories}, in: F.~Cena, A.~S. da~Silva, C.~Trattner (Eds.), Hypertext
  2014 Extended Proceedings, Vol. 1210 of {CEUR} Workshop Proceedings,
  CEUR-WS.org, 2014, pp. 1--6.
\newline\urlprefix\url{http://ceur-ws.org/Vol-1210/datawiz2014\_05.pdf}

\bibitem{Cashman2019}
D.~Cashman, S.~R. Humayoun, F.~Heimerl, K.~Park, S.~Das, J.~Thompson, B.~Saket,
  A.~Mosca, J.~Stasko, A.~Endert, M.~Gleicher, R.~Chang,
  \href{https://doi.org/10.1111/cgf.13681}{A user-based visual analytics
  workflow for exploratory model analysis}, Computer Graphics Forum 38~(3)
  (2019) 185--199.
\newblock \href {http://dx.doi.org/10.1111/cgf.13681}
  {\path{doi:10.1111/cgf.13681}}.
\newline\urlprefix\url{https://doi.org/10.1111/cgf.13681}

\bibitem{Das2019}
S.~Das, D.~Cashman, R.~Chang, A.~Endert,
  \href{https://doi.org/10.1109/mcg.2019.2922592}{{BEAMES:} {Interactive}
  multimodel steering, selection, and inspection for regression tasks}, IEEE
  Computer Graphics and Applications 39~(5) (2019) 20--32.
\newblock \href {http://dx.doi.org/10.1109/mcg.2019.2922592}
  {\path{doi:10.1109/mcg.2019.2922592}}.
\newline\urlprefix\url{https://doi.org/10.1109/mcg.2019.2922592}

\bibitem{Santos2019}
A.~Santos, S.~Castelo, C.~Felix, J.~P. Ono, B.~Yu, S.~R. Hong, C.~T. Silva,
  E.~Bertini, J.~Freire, \href{https://doi.org/10.1145/3328519.3329134}{Visus},
  in: Proceedings of the Workshop on Human-In-the-Loop Data Analytics,
  HILDA'19, ACM, New York, NY, USA, 2019, pp. 6:1--6:7.
\newblock \href {http://dx.doi.org/10.1145/3328519.3329134}
  {\path{doi:10.1145/3328519.3329134}}.
\newline\urlprefix\url{https://doi.org/10.1145/3328519.3329134}

\bibitem{Friedman2001}
J.~H. Friedman, \href{https://doi.org/10.1214/aos/1013203451}{Greedy function
  approximation: {A} gradient boosting machine.}, The Annals of Statistics
  29~(5) (2001) 1189--1232.
\newblock \href {http://dx.doi.org/10.1214/aos/1013203451}
  {\path{doi:10.1214/aos/1013203451}}.
\newline\urlprefix\url{https://doi.org/10.1214/aos/1013203451}

\bibitem{Bendre2015}
M.~Bendre, R.~Thool, V.~Thool,
  \href{https://doi.org/10.1109/ngct.2015.7375220}{Big data in precision
  agriculture: {Weather} forecasting for future farming}, in: 2015 1st
  International Conference on Next Generation Computing Technologies (NGCT),
  IEEE, 2015, pp. 744--750.
\newblock \href {http://dx.doi.org/10.1109/ngct.2015.7375220}
  {\path{doi:10.1109/ngct.2015.7375220}}.
\newline\urlprefix\url{https://doi.org/10.1109/ngct.2015.7375220}

\bibitem{malherbe2004modeling}
S.~Malherbe, V.~Fromion, N.~Hilgert, J.-M. Sablayrolles,
  \href{https://doi.org/10.1002/bit.20075}{Modeling the effects of assimilable
  nitrogen and temperature on fermentation kinetics in enological conditions},
  Biotechnology and Bioengineering 86~(3) (2004) 261--272.
\newblock \href {http://dx.doi.org/10.1002/bit.20075}
  {\path{doi:10.1002/bit.20075}}.
\newline\urlprefix\url{https://doi.org/10.1002/bit.20075}

\bibitem{papageorgiou2016fuzzy}
E.~I. Papageorgiou, K.~Kokkinos, Z.~Dikopoulou,
  \href{https://doi.org/10.1007/978-3-319-31093-0_10}{Fuzzy sets in
  agriculture}, in: Fuzzy Logic in Its 50th Year, Springer, 2016, pp. 211--233.
\newblock \href {http://dx.doi.org/10.1007/978-3-319-31093-0_10}
  {\path{doi:10.1007/978-3-319-31093-0_10}}.
\newline\urlprefix\url{https://doi.org/10.1007/978-3-319-31093-0_10}

\bibitem{munzner2009nested}
T.~Munzner, \href{https://doi.org/10.1109/tvcg.2009.111}{A nested model for
  visualization design and validation}, IEEE Transactions on Visualization and
  Computer Graphics 15~(6) (2009) 921--928.
\newblock \href {http://dx.doi.org/10.1109/tvcg.2009.111}
  {\path{doi:10.1109/tvcg.2009.111}}.
\newline\urlprefix\url{https://doi.org/10.1109/tvcg.2009.111}

\bibitem{sacha2016role}
D.~Sacha, H.~Senaratne, B.~C. Kwon, G.~Ellis, D.~A. Keim,
  \href{https://doi.org/10.1109/tvcg.2015.2467591}{The role of uncertainty,
  awareness, and trust in visual analytics}, IEEE Transactions on Visualization
  and Computer Graphics 22~(1) (2016) 240--249.
\newblock \href {http://dx.doi.org/10.1109/tvcg.2015.2467591}
  {\path{doi:10.1109/tvcg.2015.2467591}}.
\newline\urlprefix\url{https://doi.org/10.1109/tvcg.2015.2467591}

\bibitem{gutierrez2018lada}
F.~Gutiérrez, K.~Seipp, X.~Ochoa, K.~Chiluiza, T.~De~Laet, K.~Verbert,
  \href{https://doi.org/10.1016/j.chb.2018.12.004}{{LADA:} {A} learning
  analytics dashboard for academic advising}, Computers in Human Behavior 107
  (2020) 105826.
\newblock \href {http://dx.doi.org/10.1016/j.chb.2018.12.004}
  {\path{doi:10.1016/j.chb.2018.12.004}}.
\newline\urlprefix\url{https://doi.org/10.1016/j.chb.2018.12.004}

\bibitem{Kim2017}
Y.-S. Kim, K.~Reinecke, J.~Hullman,
  \href{https://doi.org/10.1145/3025453.3025592}{Explaining the gap}, in:
  Proceedings of the 2017 CHI Conference on Human Factors in Computing Systems,
  CHI '17, ACM, New York, NY, USA, 2017, pp. 1375--1386.
\newblock \href {http://dx.doi.org/10.1145/3025453.3025592}
  {\path{doi:10.1145/3025453.3025592}}.
\newline\urlprefix\url{https://doi.org/10.1145/3025453.3025592}

\bibitem{satyanarayan2017vega}
A.~Satyanarayan, D.~Moritz, K.~Wongsuphasawat, J.~Heer,
  \href{https://doi.org/10.1109/tvcg.2016.2599030}{{Vega}-{Lite}: {A} grammar
  of interactive graphics}, IEEE Transactions on Visualization and Computer
  Graphics 23~(1) (2017) 341--350.
\newblock \href {http://dx.doi.org/10.1109/tvcg.2016.2599030}
  {\path{doi:10.1109/tvcg.2016.2599030}}.
\newline\urlprefix\url{https://doi.org/10.1109/tvcg.2016.2599030}

\bibitem{satyanarayan2016reactive}
A.~Satyanarayan, R.~Russell, J.~Hoffswell, J.~Heer,
  \href{https://doi.org/10.1109/tvcg.2015.2467091}{Reactive {Vega}: {A}
  streaming dataflow architecture for declarative interactive visualization},
  IEEE Transactions on Visualization and Computer Graphics 22~(1) (2016)
  659--668.
\newblock \href {http://dx.doi.org/10.1109/tvcg.2015.2467091}
  {\path{doi:10.1109/tvcg.2015.2467091}}.
\newline\urlprefix\url{https://doi.org/10.1109/tvcg.2015.2467091}

\bibitem{Tagarakis2014}
A.~Tagarakis, S.~Koundouras, E.~Papageorgiou, Z.~Dikopoulou, S.~Fountas,
  T.~Gemtos, \href{https://doi.org/10.1007/s11119-014-9354-9}{A fuzzy inference
  system to model grape quality in vineyards}, Precision Agriculture 15~(5)
  (2014) 555--578.
\newblock \href {http://dx.doi.org/10.1007/s11119-014-9354-9}
  {\path{doi:10.1007/s11119-014-9354-9}}.
\newline\urlprefix\url{https://doi.org/10.1007/s11119-014-9354-9}

\bibitem{landry2019}
M.~Landry, A.~Bartz,
  \href{http://docs.h2o.ai/h2o/latest-stable/h2o-docs/booklets/RBooklet.pdf}{Machine
  Learning with R and {{H2O}.} Seventh Edition}, H2O.ai, Inc., 2021.
\newline\urlprefix\url{http://docs.h2o.ai/h2o/latest-stable/h2o-docs/booklets/RBooklet.pdf}

\bibitem{lecun2015deep}
Y.~LeCun, Y.~Bengio, G.~Hinton, \href{https://doi.org/10.1038/nature14539}{Deep
  learning}, Nature 521~(7553) (2015) 436--444.
\newblock \href {http://dx.doi.org/10.1038/nature14539}
  {\path{doi:10.1038/nature14539}}.
\newline\urlprefix\url{https://doi.org/10.1038/nature14539}

\bibitem{goodfellow2016deep}
I.~J. Goodfellow, Y.~Bengio, A.~C. Courville,
  \href{http://www.deeplearningbook.org/}{Deep Learning}, Adaptive computation
  and machine learning, {MIT} Press, 2016.
\newline\urlprefix\url{http://www.deeplearningbook.org/}

\bibitem{braun2006using}
V.~Braun, V.~Clarke, \href{https://doi.org/10.1191/1478088706qp063oa}{Using
  thematic analysis in psychology}, Qualitative Research in Psychology 3~(2)
  (2006) 77--101.
\newblock \href {http://dx.doi.org/10.1191/1478088706qp063oa}
  {\path{doi:10.1191/1478088706qp063oa}}.
\newline\urlprefix\url{https://doi.org/10.1191/1478088706qp063oa}

\bibitem{Ratner2019}
A.~J. Ratner, B.~Hancock, C.~R{\'{e}},
  \href{http://cidrdb.org/cidr2019/papers/p58-ratner-cidr19.pdf}{The role of
  massively multi-task and weak supervision in software 2.0}, in: 9th Biennial
  Conference on Innovative Data Systems Research, {CIDR} 2019, Asilomar, CA,
  USA, January 13-16, 2019, Online Proceedings, www.cidrdb.org, 2019, pp. 1--8.
\newline\urlprefix\url{http://cidrdb.org/cidr2019/papers/p58-ratner-cidr19.pdf}

\end{thebibliography}

\end{document}